\documentclass[journal=nalefd,manuscript=letter]{achemso}
\usepackage{amsmath}
\usepackage{bm}
\usepackage{braket}

\author{Jordi Llusar}
\affiliation{Departament de Qu\'{\i}mica F\'{\i}sica i Anal\'{\i}tica,
Universitat Jaume I, E-12080, Castell\'o de la Plana, Spain}

\author{Juan I. Climente}
\affiliation{Departament de Qu\'{\i}mica F\'{\i}sica i Anal\'{\i}tica,
Universitat Jaume I, E-12080, Castell\'o de la Plana, Spain}
\email{climente@uji.es}

\title{Charging of colloidal nanoplatelets: 
effect of Coulomb repulsion on spin and optoelectronic properties}


\begin{document}


\begin{abstract}
Colloidal semiconductor nanoplatelets combine weak lateral confinement with
strong Coulomb interactions, enhanced by dielectric confinement.
When the platelets are charged with carriers of the same sign,
this results in severe Coulomb repulsions
which shape the electronic structure. 
To illustrate this point, the shell filling of type-I (CdSe/CdS)
and type-II (CdSe/CdTe) core/crown nanoplatelets with up to 4 electrons 
or holes is investigated theoretically.
We find that Coulomb repulsions enable addition energies exceeding room 
temperature thermal energy and promote the occupation of high-spin states.
For charged excitons and biexcitons in CdSe/CdTe nanoplatelets, the repulsions further give rise to 
multi-peaked emission spectra with widely tunable (over 100 meV) energy,
and a transition from type-II to quasi-type-II band profile
as the number of electrons confined in the core increases.
We conclude that the number of excess carriers injected in nanoplatelets is a
versatile degree of freedom to modulate their magnetic and optoelectronic properties.
\end{abstract}



The number of excess carriers (electrons or holes) confined in semiconductor 
quantum dots is analogous to the charge number of ions in chemistry.
Because changes in this number are expected to modify the electronic properties,
experiments in the early years of quantum dots struggled to control it. 
Shell filling was subsequently demonstrated in gated\cite{TaruchaPRL}, 
self-assembled\cite{DrexlerPRL,ReuterPRL,EdigerNP}, carbon nanotube\cite{JorgensenNP} 
and graphene\cite{GuttingerPRL} quantum dots.
In colloidal nanocrystal quantum dots, shell filling of conduction and
valence bands was pursued using different approaches such as
scanning tunneling spectroscopy\cite{BakkersNL}, direct or remote 
doping\cite{NorrisSCI} and electrochemical injection.\cite{GuyotMA} 
The placement of resident charges in the nanocrystals translated into
substantial changes of the transport and optical properties,
including lower threshold optical gain\cite{WuNN}
and large electrochromic shifts,\cite{WangSCI,BrovelliNL}
which are of interest for lasing and sensing applications, respectively.

Recently, progress in electrochemical charge injection has reached deterministic 
and stable control of the number of confined carriers in individual nanocrystals.\cite{MorozovSCIadv}
 Charge control through doping (with impurities or surrounding molecules) 
is advancing in this direction.\cite{CapitaniNL,DuttaJPCc,DirollNL}
These achievements open path for a precise analysis of the shell structure, 
potentially unveiling few-body charge and spin interaction effects 
at a level so far restricted to fully solid state systems. 
In nanocrystal quantum dots, quantum confinement energies usually
prevail over Coulomb interaction energies.\cite{PiryatinskiNL}
 This is a similar situation to that of epitaxial quantum dots. 
One can then expect shell filling to be simply understood from the dot geometry, 
following Hund and Aufbau rules, with many-body interactions acting as a perturbation.\cite{TaruchaPRL,DrexlerPRL,NagarajaPRB,WarburtonPRB} 
A different scenario can however be foreseen for nanoplatelets (NPLs).
These are the colloidal analogous of epitaxial quantum wells, albeit with finite
and controllable lateral confinement, the possibility to develope in-plane
(core/crown) heterostructures and a strong dielectric confinement imposed
by the organic ligands.\cite{NasilowskiCR}

The flexible design and outstanding optical properties of colloidal NPLs 
have triggered intensive research over the last years aiming at applications
in optoelectronic devices.\cite{DirollJMCc,SharmaIEEE,MinNR,DuttaJPCc,VasicASS}
Much of the interest follows from the strong attraction between photogenerated
electron-hole pairs, enhanced by the quasi-2D geometry and the dielectric confinement,
which prompt large binding energies (150-250 meV) and fast radiative recombination 
rates through the so-called giant oscillator strength effect.\cite{NasilowskiCR,AchtsteinNL,RajadellPRB,NaeemPRB,ZelewskiJPCL}
One should however note that the same factors that favor strong electron-hole 
attraction, favor strong electron-electron or hole-hole repulsion too. 
In NPLs with injected carriers, these interactions ($\sim 100$ meV) 
largely exceed the small energy spacings between states of non-interacting particles, which
are set by the weak lateral confinement ($\sim 10$ meV). The resulting shell structure can then be expected to 
display non-trivial many-body phenomena, as observed e.g. in self-assembled
InAs dots charged with heavy holes.\cite{ReuterPRL,EdigerNP,HePRL} 
Possible implications of an electronic structure shaped by strong
repulsive correlations include the buildup of intrinsic ferromagnetism
and exotic spin states.\cite{HePRL,tasaki2020physics,GarciaPRL}

To investigate the effect of repulsive interactions in NPLs, in this work
we model CdSe/CdS and CdSe/CdTe core/crown heterostructures.\cite{NasilowskiCR}
%
 In these NPLs, a rectangular CdSe core is laterally surrounded by a
 CdX crown (X=S,Te). In the case of CdSe/CdS, a quasi-type-II 
 band alignment is formed, but the core lateral confinement is so weak
 that the single electron ground state stays in the core. 
 The resulting emission of neutral excitons is then like that of type-I NPLs.\cite{TessierNL}
 The CdS crown is still of practical interest inasmuch as it provides enhanced 
 absorption and higher quantum yield due to the efficient edge passivation.\cite{DirollJMCc,TessierNL,PrudnikauJACS}
 In the case of CdSe/CdTe, a type-II band alignment is formed 
 which spatially separates electrons from holes. Injected electrons accumulate in the 
CdSe core while holes do so in the peripherical crown.
This results in moderate electron-hole attractions across the interface,
which still preserve the excitonic character\cite{AntanovichNS,ScottPCCP},
and strong repulsions between electrons or holes.
The charge separation further suppresses Auger processes,
thus enabling long lived charged exciton and multiexciton species.\cite{WangCJCP}
With appropiate engineering, these structures are of interest
for light harvesting, sub-band gap emission and infrared detection,
fluorescence up-conversion systems and low-threshold lasing.\cite{MinNR,LoAM,LiACSener,KhanACS,LiACS}
  
Using effective mass Hamiltonians coupled to a configuration interaction (CI) method,
we calculate the low-lying states in NPLs with up to 4 electrons or 4 holes,
which suffice to fill the first two shells. Several phenomena of interest
are then predicted. These include (i) addition energies 
(the analogous of electron affinity in real atoms) 
over 30 meV, which imply the possibility to charge NPLs 
carrier-by-carrier at room temperature despite the weak lateral confinement;
(ii) the thermal occupation of high-spin states, enabled by
the strong Coulomb exchange energies ($20-35$ meV), which implies 
that magnetic interactions with external fields or dopants 
will be greatly enhanced in multi-electron and multi-hole systems. 
For type-II NPLs, we also calculate highly charged exciton and biexciton states. 
Further effects of charging are then observed in the emission spectrum: 
(iii) the formation of multiple peaks, which define the electronically limited
bandwidth of these systems and can be exploited for multicolor emission;
(iv) blue- and redshifts with respect to the neutral exciton 
in an energy range of over 100 meV, well beyond the typical shifts obtained 
by varying lateral confinement; and (v) the delocalization of electrons 
outside the core as repulsions exceed the conduction band offset, 
which translates into a sudden boost of the interband recombination rate,
and may be useful e.g. for the design of optical charge sensors.

\section{Results}

\begin{figure}
	\centering
	\includegraphics[width=16cm]{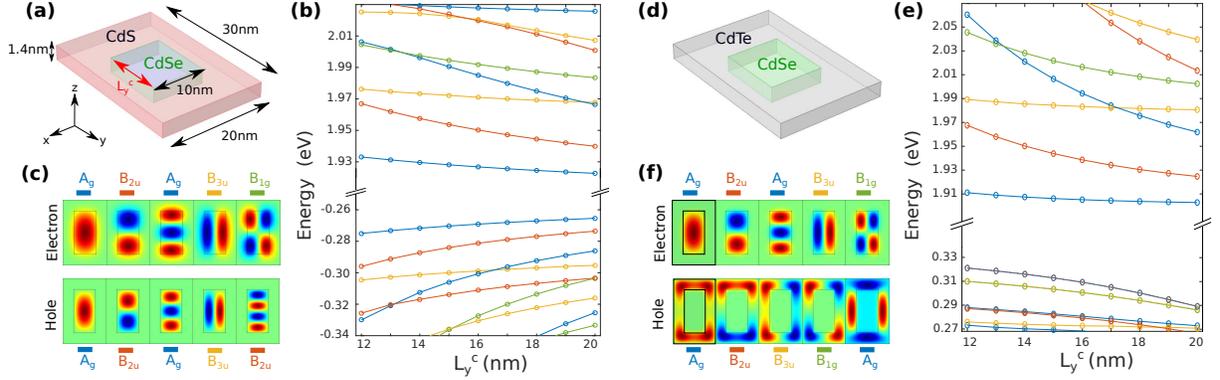}
\caption{(a) Schematic of the CdSe/CdS NPLs under study.
(b) Non-interacting electron (top) and hole (bottom) energy levels as a function of the core length.
(c) Wave functions of the lowest electron (highest hole) energy levels for $L_y^c=20$ nm.
(d-f): same but for CdSe/CdTe NPLs. The NPLs have the same dimensions as for CdSe/CdS.
All energies are referred to the top of the CdSe valence band.
The colors of the lines in (b), (c), (e) and (f) denote the irreducible 
representation of the level (within the $D_{2h}$ point group).
}\label{fig1}
\end{figure}

We consider core/crown NPLs like those shown in Fig.~\ref{fig1}a (CdSe/CdS)
and Fig.~\ref{fig1}d (CdSe/CdTe).
To investigate the role of Coulomb repulsions we will conjugate two competing
degrees of freedom: the number of confined carriers and the quantum confinement
strength.  
Thus, we take a fixed crown (lateral dimensions $20\times 30$ nm$^2$), 
and change the core size (lateral dimensions $10\times L_y^c$ nm$^2$, 
where $L_y^c=12-20$ nm is the length of the core).
 The NPL thickness is $4.5$ monolayers.
These are typical values for this kind of structure.\cite{PedettiJACS}

It is convenient to start by analyzing the energy structure of non-interacting
electrons and holes. 
 In CdSe/CdS NPLs, the first electron and hole states localize in the core (Fig.~\ref{fig1}c),
 but in CdSe/CdTe NPLs the hole moves towards the crown (Fig.~\ref{fig1}d),
 as expected from the type-I vs type-II band alignment.
 The energy dependence on the core length is similar for all
particles localized in the core.
 For instance, Fig.~\ref{fig1}b shows that increasing $L_y^c$ in CdSe/CdS NPLs
barely changes the electron or hole ground state energy ($A_g$ symmetry) 
because the lateral confinement is already weak. 
However, some of the excited states (those with nodes along the $y$ axis) 
are more sensitive to the confinement relaxation and lower their energy more, see e.g. the first
excited state ($B_{2u}$ symmetry). The same happens for electrons in CdSe/CdTe NPLs,
as can be seen in Fig.~\ref{fig1}e (top part).
 This is an indication that the NPL core is not yet in the quantum well limit,\cite{peric2021van}
and hence the density of states can be increased by making it larger.
The smaller interlevel energy spacings will translate into stronger 
electronic correlation effects, as we shall show below.

For holes in CdSe/CdTe, the behavior is different because increasing 
the core size reduces the space left in the crown.
This unstabilizes the energy levels for large $L_y^c$ values.
It is also worth noting that the top-most levels of the valence band
are formed by nearly degenerate pairs of states ($A_g$ and $B_{2u}$,
$B_{3u}$ and $B_{1g}$). This is because the core constitutes a
potential barrier which separates the crown into two symmetric sides. 
The pairs of levels are the symmetric and antisymmetric solutions
of the double box system. The top-most hole states have little
kinetic energy, so that tunneling is weak and the two solutions
are quasi degenerate.\cite{SteinmetzJPCc}


\begin{figure}
	\centering
		\includegraphics[width=12cm]{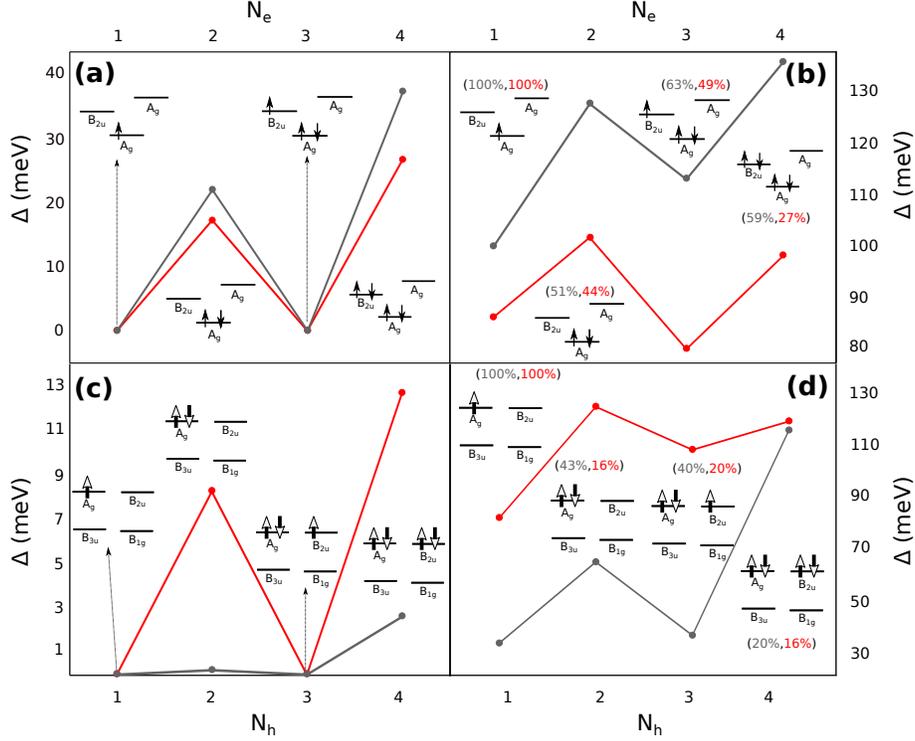}
\caption{Addition energy spectra as a function of the number of carriers
for CdSe/CdS (red line) and CdSe/CdTe (gray line) in a NPL with $L_y^c=20$ nm. 
(a) Non-interacting electrons. (b) Interacting electrons.
(c) Non-interacting holes. (d) Interacting holes.
The insets in (b) and (d) show the dominant electronic configuration 
along with its weight in the CI expansion (red value for CdSe/CdS,
grey one for CdSe/CdTe).} 
\label{fig2}
\end{figure}

We are now in a position to study the shell filling of the NPLs.
To this end, we calculate the addition energy spectrum for the two first 
conduction and valence band shells (i.e. up to 4 electrons or 4 holes).
The addition energy is the energy required to insert one additional 
charge into the nanostructure:\cite{TaruchaPRL}
\begin{equation}
\Delta = \mu(N+1)-\mu(N) = E(N+1) - 2E(N) - E(N-1).
\end{equation}
\noindent where $\mu(N)$ and $E(N)$ are the chemical potential
and ground state energy for $N$ carriers (electrons or holes).

Fig.~\ref{fig2}a shows the addition energies one would expect
for non-interacting electrons in a core with $L_y^c=20$ nm. 
In this case, a simple shell filling sequence is observed. 
Odd numbers of electrons ($N_e$) involve half-filled 
(open) shell. Because we are neglecting Coulomb interactions so far, 
introducing an extra electron in these cases requires no additional energy, 
$\Delta=0$ meV.
By contrast, even numbers of electrons involve closed shells.
Introducing an extra electron requires providing the energy
to access the next excited orbital, which is set by 
the lateral quantum confinement. This results in $\Delta \approx 20$ meV.

Upon inclusion of Coulomb interactions, major changes take place in the electron
addition spectrum, as shown in Fig.~\ref{fig2}b. Open shells have now addition 
energies exceeding 100 meV, 
which gives a direct measure of the strong electron-electron repulsions one needs to 
overcome in order to place an extra electron in the same orbital. 
Closed shells are still more stable than open ones, but Coulomb repulsions are less 
severe when the extra electron is placed in a different (orthogonal) orbital.  
Consequently, the sawtooth structure in Fig.~\ref{fig2}b is less pronounced 
than one would expect from quantum confinement alone (Fig.~\ref{fig2}a).
The latter effect is particularly remarkable in CdSe/CdS NPLs with cores under $L_y^c=15$ nm. 
In such structures, the core can only host up to two electrons. The third
electron is placed in the crown, which has much weaker repulsions.
This gives rise to a closed shell configuration ($N_e=2$) 
being less stable than an open shell one ($N_e=1$), 
see Fig.~S1 in the Supporting Information, SI.

The insets in Fig.~\ref{fig2}b show the dominant electronic configuration
for each $N_e$. One can see that the independent-particle spin-orbitals
are filled sequentially, consistent with the Aufbau principle of atoms,
but the weight of this configuration 
within the CI expansion is around 50\%. This is an indication
that strong electronic correlations (interactions beyond perturbative scheme)
are taking place in the core. 

For holes, the impact of Coulomb interactions is also drastic.
Holes in CdSe/CdS NPLs behave much like electrons in spite of the heavier masses, 
see red line in Fig.~\ref{fig2}c,d. 
Holes in CdSe/CdTe behave differently instead
because they are localized in the crown, see gray lines in the figure.
In a non-interacting picture, Fig.~\ref{fig2}c,  $\Delta \approx 0$ meV for a
number of holes $N_h=1-3$,
but $\Delta = 2-12$  meV for $N_h=4$. This reflects the valence band electronic
structure analyzed in Fig.~\ref{fig1}f. The top-most hole orbitals 
($A_g$ and $B_{2u}$) are quasi degenerate. Together with the spin degree 
of freedom, this gives a four-fold quasi-degenerate ground state. Then, adding extra
holes requires very little energy except for the closed shell, $N_h=4$.
However, when repulsions are taken into account, Fig.~\ref{fig2}d, all addition energies
shift up by at least 30 meV, and a peak emerges for $N_h=2$. The latter is 
because two holes feel comfortable sitting on opposite sides of the crown, but
adding a third hole necessarily implies stronger repulsions.

The insets of Fig.~\ref{fig2}d reveal that the weight of the main configuration in 
the CI expansion for $N_h \geq 2$ is well under $50\%$. 
This means that correlations are so strong that the Aufbau principle is no
longer a valid rule to describe holes in the NPL crown. 
Similar observations have been reported for holes in epitaxial quantum dots.\cite{EdigerNP,HePRL}

It is concluded from Fig.~\ref{fig2} that Coulomb repulsions play a dominant
role in shaping the electronic structure and addition energy spectrum of electrons and holes
in NPLs. An important practical consequence is that all $\Delta$ values exceed thermal energy at room temperature. 
This implies that electrochemical charging of NPLs\cite{DirollNC} is susceptible of being conducted electron-by-electron 
(or hole-by-hole), thus enabling deterministic control of the number of charges in spite
of the weak lateral confinement.
 The same conclusion holds for different core dimensions (Fig.S1).\\ 


The results in Figs.~\ref{fig1} and \ref{fig2} reveal that interacting electrons in 
CdSe cores experience Coulomb repulsions up to one order of magnitude 
larger than the interlevel energy spacing set by lateral confinement 
($\sim100$ meV against $\sim10$ meV for $L_y^c=20$ nm). 
This gives rise to intense electronic correlations, and
subsequently to potential conditions for the formation of 
spontaneous magnetic phases (paramagnetism and ferromagnetism).\cite{Tasaki_book,Mattis_book} 
Confirming such a point is of significant interest, 
since earlier manifestations of magnetic phases in colloidal NPLs were restricted to doping\cite{SharmaIEEE,ShornikovaACS,DavisCM,DaiCM,NajafiJPCL}
and surface-induced paramagnetism.\cite{ShornikovaNN} 
The development of intrinsic magnetism would open up new scenarios 
for spintronic and magnetic devices.
 
\begin{figure}
	\centering
		\includegraphics[width=12cm]{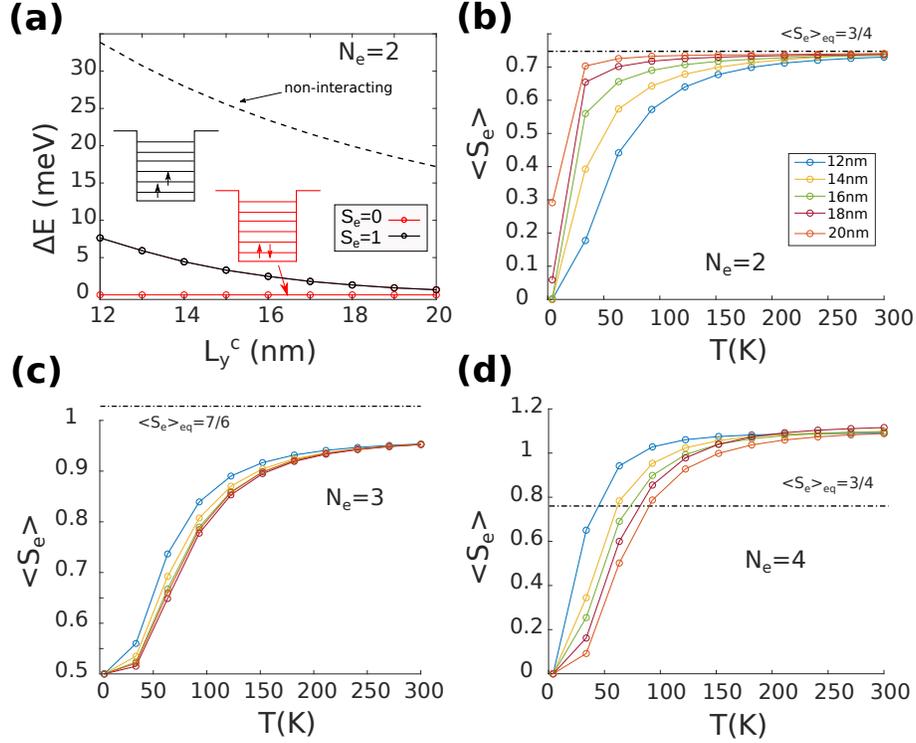}
\caption{(a) Energy splitting between ground state (singlet) 
and the first excited state (triplet) for two electrons in 
CdSe/CdS NPL with varying core length.
The insets illustrate the main electronic configuration for each state. 
The dashed line in (a) represents the triplet state of two non-interacting electrons, 
for comparison.
(b-d): expectation value of total electron spin as a function of the temperature
for different core lengths, with two (b), three (c) and four (d) electrons.
$\langle S_e \rangle_{eq}$ is the value for equal number of low and high spin states
populated.
}
\label{fig3}
\end{figure}

To explore this possibility, in Fig.~\ref{fig3} we plot the energy
difference between low and high spin states of few-electron ($N_e=2-4$) 
CdSe/CdS NPLs and the associated expectation values of the total spin quantum number
(similar results hold for CdSe/CdTe, Fig.~S2).
Fig.~\ref{fig3}a shows the energy splitting between the $N_e=2$
ground state (singlet, $S_e=0$) and the first excited state (triplet, $S_e=1$).
Clearly, as the core size increases, the triplet approaches the singlet ground state.
There are two reasons for this. One is the weakening of lateral confinement, which
selectively relaxes the first excited orbital ($B_{2u}$, $p_y$-like) but not the
lowest one ($A_g$, $s$-like), as discussed before in Fig.~\ref{fig1}.
The other reason is that strong Coulomb repulsions imply large Coulomb
exchange energies as well, which stabilize triplets as compared to singlets. 
Fig.~\ref{fig3}a compares the triplet energy for two non-interacting 
(dashed line) and two interacting (solid line with black circles) electrons.
The energy decrease for non-interacting electrons is merely due to the weakened confinement,
while that of the interacting case further benefits from exchange energies as large
as $20-30$ meV. 
Consequently, for two interacting electrons, the $S_e=1$ state is only 8 meV away 
from the $S_e=0$ ground state when $L_y^c=12$ nm, and is nearly degenerate when $L_y^c=20$ nm.
It follows that thermal occupation of high spin states is feasible at room
temperature or even earlier. 
As shown in Fig.~\ref{fig3}b, with increasing temperature the total spin 
$\langle S_e \rangle$ rapidly departs from $\langle S_e \rangle=0$ (pure singlet), 
which is the value one would obtain in strongly confined nanocrystals, 
and reaches $\langle S_e \rangle \approx 3/4$, which implies equal population
of singlet and triplet states. For large cores ($L_y^c=20$ nm), this is
achieved at temperatures under 100K. 
The practical implication is that spontaneous paramagnetic response 
must be expected for usual sizes of CdSe cores except at the lowest 
temperatures. 

The ground state of a $N_e=2$ system in the absence of external
magnetic fields or spin-orbit interaction is always spin-singlet.\cite{Mattis_book}
In Fig.~\ref{fig3}c,d we explore whether this situation can
be reversed for $N_e=3-4$ NPLs, and high-spin ground states show up.
For the geometries we address the answer is negative, 
but again high-spin states are occupied at room
temperature. In fact, for $N_e=4$ and 300K we obtain $\langle S_e \rangle > 3/4$,
see Fig.~\ref{fig3}d. That is, thermal population of triplet states exceeds
that of singlet ones. The reason is that more than one triplet state becomes 
reachable at room temperature, see Fig.~S2. 

We conclude from Fig.~\ref{fig3} that the large Coulomb exchange energy
and the weak lateral confinement of colloidal NPLs enable the occupation of high spin states
from low temperatures, which should provide few-electron NPLs with paramagnetic 
behavior and the possibility to form embryonic ferromagnetic phases. 
Because the total spin of electrons couples to that of magnetic dopants
and to external fields,\cite{SharmaIEEE,ShornikovaNN,GovorovCRP}
these findings anticipate that charged NPLs will display enhanced magnetic response
as compared to NPLs with conventional photoexcitation of neutral excitons.

\begin{figure}
	\centering
		\includegraphics[width=16cm]{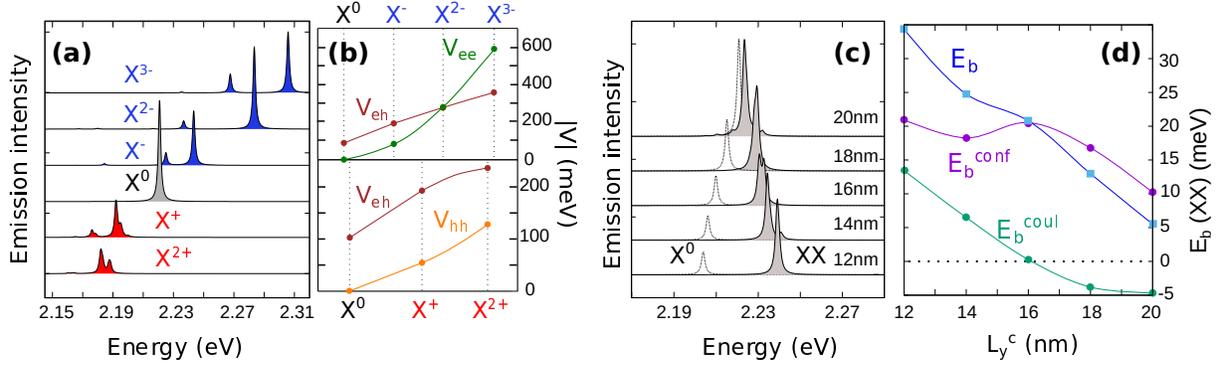}
\caption{(a) Emission spectrum of neutral and charged excitons in a CdSe/CdTe NPL.
	(b) Corresponding mean value of Coulomb attractions and repulsions.
	Coulomb terms are depicted in absolute value.
	(c) Comparison between the emission spectrum of biexcitons and excitons
	for different core lengths, $L_y^c$.
	(d) Corresponding biexciton binding energy, along with its kinetic and Coulomb contributions (see text).
	In (a) and (b), $L_y^c=20$ nm. The spectra are simulated at $20$ K.}
\label{fig4}
\end{figure}

Next, we go beyond spin effects in few-fermion systems and
study the influence of charging on the optoelectronic properties.
To this end we focus on type-II CdSe/CdTe NPLs, where repulsions are
not balanced by attractions and hence lead to more conspicuous effects.
 Figure \ref{fig4}a shows the calculated emission spectrum for 
a NPL with a $L_y^c=20$ nm core at $T=20$ K, for different excitonic complexes.
$X^0$ stands for a neutral exciton (one interacting electron-hole pair),
which is the usual dominating species in photoexcited CdSe/CdTe NPLs.\cite{SteinmetzJPCc,PandyaACS}
When extra charges are introduced in the system, $X^{n \pm}$ complexes are formed,
where $n$ is the number of charges (positive or negative) added to the neutral exciton.
Interestingly, varying the number of charges in the NPL has a substantial
effect on the optical spectrum. Negatively charged excitons become increasingly blueshifted,
while positively charged excitons are redshifted instead. In both cases, charging leads
to multi-peaked emission. 

The energetic shift between $X^{2+}$ and $X^{3-}$, over 100 meV, is of considerable magnitude.
It is larger than the spectral shifts one can obtain in colloidal NPLs by modifying the weak 
lateral confiment (few tens of meV).\cite{BertrandCC,DiGiacomoCM} 
It is also much larger that typical shifts observed upon charging of type-I quantum dots
and NPLs.\cite{EdigerNP,AyariNS,PengPRM,Macias_arxiv} 
Besides, its origin cannot be simply ascribed to population of higher excited orbitals,
as one would expect when bleaching quantum dots, because the energy spacing between consecutive
orbitals in NPLs is small (Fig.~\ref{fig1}).
The shift arises mostly from the unbalanced repulsive and attractive Coulomb interactions. 
 To gain qualitative understanding, 
we compare the relative strength of attractions and repulsions for each excitonic complex.
We calculate the expectation value of the terms contributing to the total energy of $X^{n \pm}$
:
\begin{equation}
\langle E_{tot} \rangle = 
 \langle E_e \rangle + 
 \langle E_h \rangle + 
 \langle V_{eh} \rangle + 
 \langle V_{ee} \rangle + 
 \langle V_{hh} \rangle. 
 \label{eq:Etot}
\end{equation}
\noindent Here $E_{tot}$ is the total energy of the $X^{n \pm}$ ground state, 
$E_e$ ($E_h$) is the sum of the energies of the non-interacting electrons (holes) forming the complex,
$V_{eh}$ is the sum of the attractions between electron-hole pairs and $V_{ee}$ ($V_{hh}$) that of the 
electron-electron (hole-hole) repulsions.
Figure \ref{fig4}b compares the absolute value of repulsions and attractions for $X^{n-}$ 
(top panel) and $X^{n+}$ (bottom panel).  In both cases, the attractions $\langle V_{eh} \rangle$ 
increase (in absolute value) with the number of carriers, because the electron finds more holes to 
interact with (or vice-versa) across the CdSe/CdTe interface. 
The repulsions show however contrasting behaviors for electrons and holes. 
$\langle V_{ee} \rangle$ shows a rapid --superlinear-- increase, 
reflecting the strong electron-electron interactions within the CdSe core,
and leads to repulsions surpassing attractions in $X^{n-}$.
By contrast, holes are localized in the CdTe crown, with a large core separating the two symmetric sides.
This yields relatively weak repulsions, $ \langle V_{hh} \rangle $, which remain smaller than attractions.
In short, in $X^{n-}$ repulsions prevail over attractions, while in $X^{n+}$ the opposite holds.
This is directly connected to the blueshift (redshift) observed in Fig.~\ref{fig4}a.
It follows from Fig.~\ref{fig4}a,b that Coulomb interactions permit using 
the number of extra charges injected in type-II NPLs as a tool for broad 
and reversible tuning of the emission wave lengths around the value set 
by the NPL thickness. The same finding holds at room temperature, see Fig.~S4.
Experimental demonstration of Coulomb induced shifts upon chemical charging
have been reported in CdSe/CdTe quantum dot nanocrystals.\cite{BangJPCc}
Our results indicate that NPLs, being in the weak (lateral) confinement regime,
can also benefit from this effect. 
 
As for the multi-peaked emission of charged species in Fig.~\ref{fig4}a, it arises from
transitions involving not only the band edge states, but also high-spin states at low energy\cite{SteinmetzJPCc}, 
with small satellites further arising from shake-up processes.\cite{AntolinezNL,LlusarACSph}
A detailed spectral assignment of the trion states ($X^-$ and $X^+$) is provided in the SI, Figs.~S5 and S6.

In CdSe/CdTe NPLs, the slow Auger relaxation enables the formation not only of charged excitons but also
of long-lived biexcitons ($XX$).\cite{WangCJCP}
Biexcitons in type-II heterostructures are of interest for applications such as single-exciton lasing,
since repulsive exciton-exciton interactions lift the balance between absorption and stimulated emission.\cite{KlimovNAT,PiryatinskiNL}
To give insight into the possibilities of engineering such interactions through changes in the dimensions of CdSe/CdTe NPLs,
in Fig.~\ref{fig4}c we plot the exciton and biexciton emission spectrum for NPLs with different core length.
As can be seen, by varying $L_y^c$, the shift between $XX$ and $X^0$ peaks is
tuned from $\sim 40$ meV ($L_y^c=12$ nm) to nearly zero ($L_y^c=20$ nm). 
Again, this magnitude is similar to that reported for core/shell nanocrystals,\cite{PiryatinskiNL}
but here it is achieved within the weak confinement regime,
which permits keeping associated physical phenomena, such as reduced Auger relaxation,\cite{KunnemanJPCL}
giant oscillator strength effect,\cite{ScottPCCP} or the enhanced magnetic moment we predict in Fig.~\ref{fig3}.
The tunability of the spectral shift holds at room temperature as well, see Fig.~S7.
 
The physical origin of the shift is also different from that of nanocrystals, as we explain next.
In both cases the emission peak is mainly related to the band edge transition 
(see Fig.~S8 for a spectral assignment of the $XX$ spectrum).
Thus, the spectral shift between $XX$ and $X^0$ corresponds to the biexciton binding energy,
$E_b (XX) = E_{tot}(XX) - 2E_{tot}(X)$ (here positive sign means unbound biexciton).
Using expectation values, Eq.~(\ref{eq:Etot}), $E_b(XX)$ can be decomposed as:
\begin{equation}
\langle E_b (XX) \rangle = \langle E^{conf} \rangle + \langle E^{coul} \rangle.
\end{equation}
\noindent Here $E^{conf}= (E_e(XX)-2E_e(X^0)) + (E_h(XX)-2E_h(X^0))$ is the spatial confinement contribution
to the binding energy, associated to changes in the energy of occupied single-particle spin-orbitals.
In turn, $E^{coul}=V_{eh}(XX)+V_{ee}(XX)+V_{hh}(XX)-2V_{eh}(X^0)$ is the Coulomb contribution,
associated to changes in the relative strength of Coulomb interactions. 
In strongly confined quantum dots and nanocrystals, a perturbative description of Coulomb interations applies,
where both $XX$ and $X^0$ have electrons (holes) in the $1S_e$ ($1S_h$) orbital 
(first $A_g$ orbital in Fig.~\ref{fig1}e).
Then, $E^{conf} \approx 0$ and the biexciton shift is well explained from changes in 
$E^{coul}$ alone.\cite{PiryatinskiNL,DalgarnoPRB}
However, NPLs are in a strongly correlated regime, with Coulomb repulsions
promoting the occupation of excited orbitals (recall the configuration mixing
in Fig.~\ref{fig2}b,d). Consequently, $E^{conf}$ gives a major contribution to $E_b(XX)$.
This can be seen in Fig.~\ref{fig4}d. Even at $L_y^c=16$ nm, when $E^{coul} \approx 0$ 
(attractions equal repulsions), $E_b(XX) \approx E^{conf} \approx 20$ meV, 
which explains the net blueshift of this core size in Fig.~\ref{fig4}(c).\\

\begin{figure}
	\centering
		\includegraphics[width=14cm]{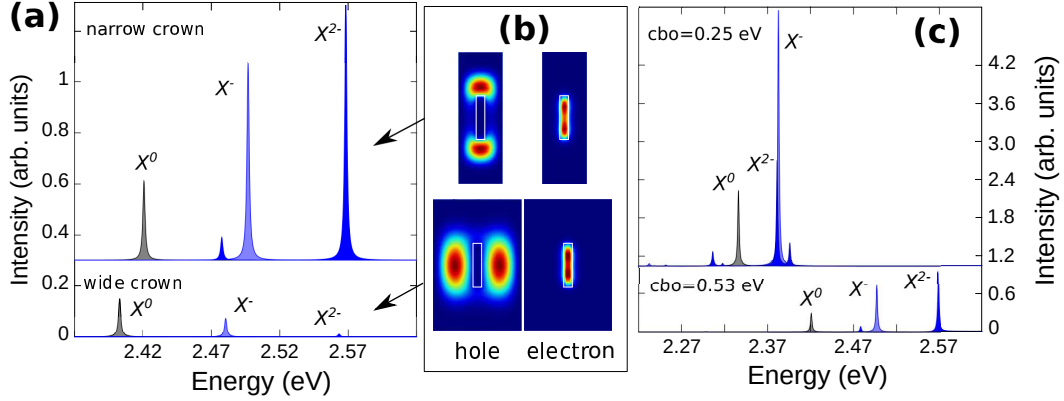}
\caption{(a) Emission spectrum of $X^0$, $X^-$ and $X^{2-}$ for NPLs
with wide ($20 \times 30$ nm$^2$) and narrow ($10 \times 30$ nm$^2$) crown.
(b) Corresponding hole (left) and two-electron (right) charge densities for $X^-$.
(c) Emission spectrum of the narrow crown NPL, now comparing conduction band
offset $0.53$ eV (as in CdSe/CdTe) with $0.25$ eV (as in CdSe/CdSe$_{0.5}$Te$_{0.5}$).
The spectra are simulated at 4K and are offset vertically for clarity.} 
\label{fig5}
\end{figure}

We have so far shown the potential of Coulomb repulsions to modulate the
emission energy of charged excitons and biexcitons in type-II NPLs. 
The last question we address is whether a similar modulation can be 
achieved on the emission rate. In type-I NPLs, trions ($X^\pm$)
have smaller oscillator strength than $X^0$,\cite{AyariNS} because 
the giant oscillator strength effect is diluted.\cite{Macias_arxiv}
In Fig.~\ref{fig4}a, however, no major change is observed between $X^0$
and $X^{n \pm}$, which suggests that in type-II NPLs the reduced 
electron-hole overlap gives rise to a distinct behavior. 
Nonetheless, substantial modulation of the oscillator strength can be 
obtained if one uses a proper structural design, in which Coulomb interactions
are exploited to force a transition from type-II to quasi-type-II
localization of charged excitons.
To illustrate this effect, we consider two NPLs with the same core 
($2 \times 10$ nm$^2$) but different crown dimensions:
$20\times 30$ nm$^2$ 
vs $10\times 30$ nm$^2$. 
Fig.~\ref{fig5}a compares the resulting emission spectrum 
for $X^0$, $X^-$ and $X^{2-}$ in both NPLs.
Clearly, the narrow-crown NPL shows a gradual enhancement of the
intensity as the number of electrons increases, 
which is not observed in the wide-crown NPL.
The different response is related to the localization of holes in each crown.
As shown in Fig.~\ref{fig5}b, in a wide crown the hole charge density
exhibits transversal localization, while in a narrow crown lateral confinement
favors longitudinal localization. In turn, the electron charge density is similar in 
both geometries. As the number of electrons increases, Coulomb repulsions stretch
the electron density along the longitudinal axis (see electron panels in Fig.~\ref{fig5}b,
which correspond to $X^-$). This enhances leakage into the crown, which results
in stronger electron-hole overlap --and hence greater emission intensity-- 
in the narrow crown configuration only.

Further enhancement of the intensity can be obtained by reducing 
the conduction band offset, e.g. by using alloyed (CdSeTe) crowns.\cite{DufourJPCc,KelestemurJPCc2}
To estimate the magnitude of this effect, in Fig.~\ref{fig5}c
we compare the emission of the narrow CdSe/CdTe NPL 
to that of the same NPL with halved conduction band-offset (close to that of CdSe$_{0.5}$Te$_{0.5}$). 
One can see that in the latter case, a greater increase of the emission 
intensity is obtained when the number of electrons becomes larger (cf. $X^0$ and $X^-$).
This is because electron-electron repulsions find it easier to overcome
the crown potential barrier, and increase the leakage outside the core.
 
All in all, Fig.~\ref{fig5} evidences that the emission intensity of CdSe/CdTe NPLs 
can be modulated through the number of injected electrons in the core. With appropiate 
structural design, Coulomb interactions make excitonic species gradually reduce their 
indirect character.
Electron-electron repulsions play an important role to this regard by prompting electron 
delocalization outside the core, but it is worth noting that so do electron-hole attractions
(see Fig.~S9).
This carrier-sensitive optical response could be interesting for ratiometric probing of charges in the NPL.

\section{Conclusions}

In conclusion, we have shown that the number of carriers confined in 
type-I and type-II colloidal NPLs is a powerful tool to control its electronic structure. 
Several phenomena can be observed in the spin and optoelectronic physics,
which include the formation of intrinsic paramagnetic configurations,
large energy shifts in the optical spectrum (to either red or blue) 
and changes in the radiative recombination probability.
These effects are characteristic of colloidal NPLs, 
as they are mostly driven by Coulomb repulsions
--which are small in bulk-- and require weak lateral 
confinement --which is missing in quantum dots--.
Proper engineering of the band structure through the number of 
carriers opens paths for applications such as diluted magnetic
semiconductors with enhanced electron-mediated ferromagnetism,\cite{FernandezPRL,QuPRL}
single-exciton lasing, voltage-tunable emission wavelength devices
or optical sensing of charges.



\section{Methods}

Calculations are carried within k$\cdot$p theory framework. 
Non-interacting electron and hole states are calculated with single-band Hamiltonians
including core/crown lattice mismatch strain in a continuum elastic model\cite{SteinmetzJPCc} 
and self-energy corrections to account for dielectric mismatch.\cite{MovillaJPCL} 
Low temperature band gaps of CdSe ($1.76$ eV) and CdTe ($1.6$ eV) are taken
from Ref.~\citenum{Sadao_book} and the rest of material parameters from 
Ref.~\citenum{LlusarJPCc}.
 
 Many-body eigenstates and eigenenergies are calculated within a 
 full CI method, using \emph{CItool} codes.\cite{citool}
 Coulomb integrals for the CI matrix elements, 
including the enhancement coming from dielectric confinement, 
are calculated solving the Poisson equation with Comsol 4.2.
 The CI basis set for CdSe/CdS NPLs is formed by all possible combinations 
 of the first 28 independent-electron and 24 independent-hole spin-orbitals
 (20 and 24 for CdSe/CdTe).

 Charged exciton and biexciton configurations are then defined all
possible Hartree products between the few-electron and few-hole Slater determinants,
consistent with spin and symmetry requirements.
Optical spectra are calculated within the dipole approximation,\cite{Pawel_book}
assuming Lorentzian bands with linewidth of 1 meV.

{\bf Supporting Information} 

This material is available free of charge via the internet at http://pubs.acs.org.

Additional calculations on core size dependence of addition energies, 
spin expectation value for CdSe/CdTe NPLs, emission spectra at high temperature,
spectral assignments for trions and biexciton, and effect of electron-hole
attraction on the emission of CdSe/CdTe NPLs are given.

\begin{acknowledgement}
The authors acknowledge support from MICINN project CTQ2017-83781-P.
We are grateful to Iwan Moreels, Thierry Barisien and Josep Planelles
for discussions.
\end{acknowledgement}

\bibliography{chargebib}

\newpage

{\large {\bf For Table of Contents Use Only}}


\begin{tocentry}
	\begin{center}
\includegraphics{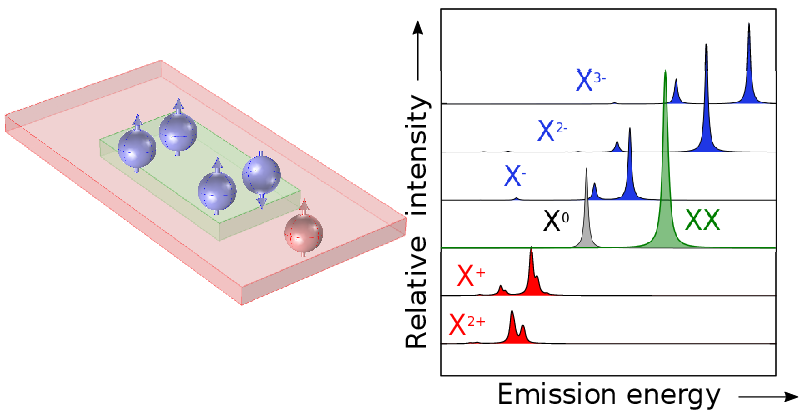}
	\end{center}
\end{tocentry}

Left: schematic of NPL with few interacting electrons (with high spin configuration) and hole
in a type-II structure. 
Right: low temperature emission spectrum as a function of the number of charges.


\end{document}